

\def\SDE{Schwinger--Dyson equation}
\def\RGE{renormalization group equation}

\def\OPI{one particle irreducible}
\def\OPR{one particle reducible}
\def\rhs{right hand side}
\def\lhs{left hand side}
\def\pslash{p\!\!/}
\def\kslash{k\!\!/}

\def\p2inf{\mathrel{\mathop{\sim}\limits_{\scriptscriptstyle
{p^2 \to \infty }}}}
\def\kap2inf{\mathrel{\mathop{\sim}\limits_{\scriptscriptstyle
{\kappa \to \infty }}}}
\def\x2inf{\mathrel{\mathop{\sim}\limits_{\scriptscriptstyle
{x \to \infty }}}}
\def\Lam2inf{\mathrel{\mathop{\sim}\limits_{\scriptscriptstyle
{\Lambda \to \infty }}}}
\def\twiddles#1{\mathrel{\mathop{\sim}\limits_
                        {\scriptscriptstyle {#1\to \infty }}}}
\def\frac#1#2{{{#1}\over {#2}}}
\def\half{\hbox{${1\over 2}$}}
\def\third{\hbox{${1\over 3}$}}
\def\quarter{\hbox{${1\over 4}$}}
\def\smallfrac#1#2{\hbox{${#1\over #2}$}}

\def\Mev{{\rm MeV}}
\def\Gev{{\rm GeV}}

\def\matele#1#2#3{\langle {#1} \vert {#2} \vert {#3} \rangle }

\def\vpy#1#2#3{\hbox{{\bf #1} (#3) #2}}
\catcode`@=11 
\def\slash#1{\mathord{\mathpalette\c@ncel#1}}
 \def\c@ncel#1#2{\ooalign{$\hfil#1\mkern1mu/\hfil$\crcr$#1#2$}}
\def\lsim{\mathrel{\mathpalette\@versim<}}
\def\gsim{\mathrel{\mathpalette\@versim>}}
 \def\@versim#1#2{\lower0.2ex\vbox{\baselineskip\z@skip\lineskip\z@skip
       \lineskiplimit\z@\ialign{$\m@th#1\hfil##$\crcr#2\crcr\sim\crcr}}}
\catcode`@=12 
\input harvmac
\pageno=0
\tolerance=10000
\hfuzz=5pt
\baselineskip 12pt plus 2pt minus 2pt
\line{\hfil   DFTT 92/69}
\line{\hfil   OUTP-92-35P}
\line{\hfil May 1993}
\medskip
\centerline{\bf A DYNAMICAL RESOLUTION OF}
\centerline{\bf THE SIGMA TERM PUZZLE}
\vskip 36pt\centerline{Richard
D.~Ball$^{a}$\footnote{${}^\dagger$}{Address from October 1993: Theory
Division, CERN, CH-1211 Gen\`eve 23, Switzerland.},
Stefano Forte$^{b}{}^\dagger$, and Jason Tigg$^{a}$}
\vskip 12pt
\centerline{\it Theoretical Physics, 1 Keble Road,}
\centerline{\it  Oxford OX1 3NP, U.K.~$^{a}$}
\vskip 10pt
\centerline{\it and }
\vskip 10pt
\centerline{\it I.N.F.N., Sezione di Torino}
\centerline{\it via P.~Giuria 1, I-10125 Torino, Italy.$^{b}$}
\vskip 1.in
{\narrower\baselineskip 10pt
\centerline{\bf Abstract}
\medskip
We propose a resolution of the puzzle posed by the discrepancy between the
value of the pion--nucleon sigma term inferred from pion--nucleon
scattering, and that deduced from baryon mass splittings using the
Zweig rule. We show that there is a significant hypercharge--dependent
dynamical contribution to baryon masses, not hitherto included in the
analysis, which may be estimated using the scale Ward identity,
and computed by solution of the Schwinger--Dyson
equation for the quark self--energy.
We find that the discrepancy is completely resolved
without any need for Zweig rule violation.
\vskip 10pt
}

\vskip .7in
\centerline{Submitted to: {\it Nuclear Physics B}}
\vfill
\eject
\noblackbox
\newsec{The Sigma Term Puzzle}
The  sigma term puzzle \ref\jaf{See e.g. R.~L.~Jaffe and C.~L.~Korpa, {\it
Comm. Nucl. Part. Phys.}, \vpy{17}{163}{1987}.}
is one of several instances in which experimental
data seem to disagree with our understanding of the
structure of the nucleon based
on naive quark model intuition. The pion--nucleon sigma term is
defined as the nucleon matrix element of the light quark mass term in
the QCD Hamiltonian:
\eqn\sigmadef
{\sigma=\bar{m}\langle N |\left( \bar u u +  \bar d d\right)|N\rangle,}
where, ignoring isospin breaking, $\bar m={1\over 2} (m_u+m_d)$.
It is physically interesting because it relates the pattern of
chiral symmetry breaking in QCD to the quark content of the nucleon
\jaf\ref\don{J.~F.~Donoghue and C.~R.~Nappi,
{\it Phys Lett.} \vpy{168B}{105}{1986}.}.

The value of $\sigma$
can be determined experimentally by using current algebra to
connect the matrix element \sigmadef\ to
the value of the isospin even pion--nucleon scattering amplitude at
zero momentum transfer
\ref\cheng{T.-P.~Cheng and R.~Dashen, {\it Phys. Rev. Lett.}
\vpy{26}{594}{1971}.}. Taking several subtleties in the
extraction of the matrix element from
the scattering data into account\ref\gaslesa{J.~Gasser, H.~Leutwyler
and M.~E.~Sainio, {\it Phys. Lett.} \vpy{B253}{252}{1991}.} leads to the
value $\sigma\approx 45\;\Mev $, with an overall experimental and theoretical
uncertainty of perhaps up to $10\;\Mev $.

On the other hand, the
sigma term may also be related to the mass splittings in the baryon octet
\ref\chli{T.-P.~Cheng, {\it Phys. Rev.} \vpy{D13}{2161}{1976}\semi
see also T.-P.~Cheng and L.-F.~Li, ``Gauge theory of
elementary particle physics'' (Oxford U.P., Oxford, 1984).}. If we
assume that there are no strange quarks in the
nucleon (according to the Zweig rule), or more specifically
that $\langle N|\bar s s |N\rangle=0$,
then the sigma term \sigmadef\ is equal to the octet sigma term
\eqn\sigmoc
{\sigma_8=\bar m\langle N |
\left(\bar u u+\bar d d -2 \bar s s\right)|N\rangle ,}
which in turn is proportional to the nucleon matrix element
of the octet portion ${\cal H}_8$ of the quark mass term in the QCD
Hamiltonian:
\eqn\hamoc
{\sigma_8=\frac{3}{1 - m_s/\bar{m}}\langle N |{\cal H}_8|N\rangle}
where
\eqn\ocbr
{ {\cal H}_8=\third\left(\bar m- m_s\right)
\left(\bar{u}u+\bar{d}d-2\bar{s}s\right).}

However, in the quark model the mass splittings are assumed, using first order
perturbation theory, to be due to the
SU(3) octet component of the (effective) strong Hamiltonian
${\cal O}_8$; this leads to the relation
\eqn\bmass
{M_B=M_0+M_1 \langle B|Y|B\rangle +
M_2\langle B|\big(I(I+1)-\quarter Y^2\big)|B\rangle,}
where $I$ is the isospin operator, and $Y$ the hypercharge.
Eq.\bmass, which expresses the masses of all the octet baryons in terms of two
parameters, is in excellent agreement with the data.
The matrix element of the symmetry breaking operator ${\cal O}_8$
between nucleon states can then be determined by SU(3) algebra;
\eqn\masssplit
{\eqalign{\langle N |{\cal O}_8|N\rangle &= M_1 - \half M_2 \cr
&= M_\Lambda -
M_\Xi=\smallfrac{2}{3}M_N-\third\left(M_{\Xi}+M_{\Sigma}\right).\cr}}
Now, if ${\cal O}_8$ is
identified with the octet component of the QCD Hamiltonian,
namely ${\cal H}_8$, using the value of the
ratio $m_s/\bar{m}=25\pm 5$ determined through current
algebra \ref\mas{J.~Gasser and H.~Leutwyler,
{\it Phys. Rep.}, \vpy{87}{77}{1982}.}
\ref\DHW{J.~F.~Donoghue, B.~R.~Holstein and D.~Wyler, {\it Phys. Rev.
Lett.}, \vpy{69}{3444}{1992}.} eq.\hamoc\ determines
$\sigma _8=25\;\Mev $, with an uncertainty of about $5\;\Mev $.
The large percentage discrepancy between the values of $\sigma$
and $\sigma _8$ is known as the sigma term puzzle.

It has now become rather fashionable to see this discrepancy as
a consequence of the failure of the assumption upon which
the identification of $\sigma$ and $\sigma_8$ is based,
namely the Zweig rule; one then infers that the strange matrix element
$\matele{N}{\bar{s}s}{N}$ is relatively large.
More than $300\;\Mev$ of the
nucleon mass would then be due to the strange quark condensate, and
kaons could condense out in nuclear matter at low densities
\ref\KaNe{D.~B.~Kaplan and A.~Nelson, {\it Phys. Lett.} \vpy{175B}{57}{1986}.}.
Alternatively, the discrepancy could be interpreted as a breakdown of
first-order perturbation theory, i.e., of the linear dependence of the mass
splittings on the symmetry breaking operator which is
used to derive eq.\bmass\
and relate the mass splittings to the matrix element \hamoc.

Both of these alternatives are rather unpalatable,
since they are extremely hard to reconcile with the successfulness
of the quark model mass formulae which are
derived using the same assumptions.
For example, the Gell-Mann--Okubo mass formula
\masssplit\ is accurate to a few percent.
Also, Weinberg has determined  the current quark
masses\ref\Weinberg{S.~Weinberg,
in ``A Festschrift for I.I.~Rabi'',
ed.~L.~Motz (New York Acad. Sci., New York, 1977).}
assuming again a linear dependence of baryon masses on quark masses;
assuming that the operator $\bar\psi\psi$ is
proportional to the quark plus antiquark number,
the mass splittings are given by the difference in quark content
of the various hadrons. This is
essentially equivalent to
eq.\bmass\ with $M_2=0$, and is phenomenologically accurate to
about 20\% \ref\pbppap{M.~Anselmino and S.~Forte,
Torino preprint DFTT 92/6 (1992).}\ref\got{S.~Forte,
{\it Phys. Rev.} \vpy{D47}{1842}{1993}.} in agreement with the fact that
fitting eq.\bmass\ to the baryon octet spectrum leads to $M_2\simeq
\smallfrac{1}{5}M_1$.

Here we will show that there is a third possibility
which successfully resolves the puzzle; namely that whereas the mass
splittings are indeed given by a linear mass formula according to eq.\bmass,
the identification of matrix elements of
${\cal O}_8$ with those of ${\cal H}_8$ is incorrect due to the fact that
most of the baryon mass, and thus a substantial proportion of the mass
splittings, arises dynamically from the trace anomaly.
Once this contribution is taken into account, it follows that
the octet operator responsible for mass splittings
differs from ${\cal H}_8$
due to the presence of an additional term, generated dynamically,
which may be viewed as a nonsinglet gluonic contribution to the mass
splittings. Thus the puzzle is resolved, but
the quark model results, which follow from the assumption of linear
dependence on an octet operator, are preserved.

In section II we show that the conformal anomaly equation leads to
an exact relation between the sigma term and mass splittings, which
implies that the identification of the nonsinglet matrix elements
of ${\cal O}_8$ and
${\cal H}_8$ is indeed spoiled by the presence of a
dynamical contribution to the
masses. We then use a Ward identity to argue that this contribution is
nonsinglet and large enough to account for the observed discrepancy. In
section III we test this explanation by attempting a computation of the
nonsinglet part of the dynamically generated mass,
solving the \SDE\ for the quark self--energy in
various approximations.
We find that this gives a value for the sigma term perfectly
consistent with the data, despite large uncertainties in the infrared
dynamics responsible for the mass generation.
Conclusions are drawn in section IV.

\newsec{The Trace Anomaly and Dynamical Mass Generation}

In order to discuss the relationship of the mass splittings to the matrix
elements of
${\cal H}_8$ we exploit an {\it exact} relation between
the matrix elements of the sigma term and the nucleon masses.
Classically the sigma term \sigmadef\ is equal to the
divergence of the  Noether current for scale transformations,
which in turn equals the trace of the energy-momentum tensor.
In the quantized theory, this implies a Ward identity
that allows us to relate the matrix elements of the trace of the
energy-momentum tensor to the masses of physical states.

\subsec{The Scale Ward Identity}

Consider the dilation current
\ref\curr{See e.g. V.~De~Alfaro, S.~Fubini, G.~Furlan and C.~Rossetti,
``Currents in Hadron Physics'' (North--Holland, Amsterdam, 1973).}
$j^\mu_D=x_\nu T^{\mu\nu}$. This is the
Noether current for scale transformations; its
divergence is equal to the trace of
the energy-momentum tensor $T^{\mu\nu}$.
In the quantized theory, the trace of the energy momentum tensor
satisfies on--shell the operator equation \ref\col{J.~C.~Collins, A.~Duncan and
S.~Joglekar, {\it Phys. Rev.} \vpy{D16}{438}{1977}\semi
N.~K.~Nielsen, {\it Nucl. Phys.} \vpy{B120}{212}{1977}.}
\def\sferm{\hbox{$\sum_i$} m_i\bar\psi_i\psi_i}
\def\sglue{\hbox{${{\beta(\alpha_s)}\over{4\alpha_s}}$}
G^{\mu\nu}_aG_{\mu\nu}^a}
\eqn\noeth
{\partial_\mu j^\mu_D=T^\mu{}_\mu=(1+\gamma_m)\sferm +\sglue ,}
where the sum runs over all quark flavors,  $G^{\mu\nu}_a$ is the gluon field
strength,
$\beta\equiv\frac{d\alpha_s(\mu )}{d\ln\mu }$ is the beta-function
for the strong coupling $\alpha_s\equiv\frac{g^2}{4\pi}$, and
$\gamma_m(\mu )\equiv -\frac{d\ln m(\mu )}{d\ln\mu}$
is the mass anomalous dimension.  All operators
appearing in eq.\noeth\ are renormalized and normal--ordered.
The last term  on the \rhs\ of eq.\noeth\ is
due to the conformal anomaly;
the anomalous dimension $\gamma_m$ is present because
the anomaly term and the mass term in eq.\noeth\ are not separately
scale invariant (i.e., they mix upon renormalization),
while the energy momentum tensor is
(up to surface terms) \col. Taking matrix elements of eq.\noeth\
generates the scale Ward identities of QCD.

On the other hand, the forward matrix element of
the trace of $T^{\mu\nu}$ between hadron states is just the mass of
the hadron,
due to Lorentz invariance and the absence of a massless
scalar Goldstone boson \curr. It follows that
the  baryon matrix element of eq.\noeth\ is
\eqn\ward
{\langle B|\left(1+\gamma_m\right)\sferm +\sglue |B\rangle=M_B,}
which is the desired Ward identity.

Eq.\ward\ shows that the first-order perturbative expression
which equates mass splittings to nonsinglet matrix
elements of the mass term is protected by the scale Ward
identity, and exact up to quantum corrections. These appear in eq.\ward\
in two distinct instances, namely, the nonzero value of the anomalous dimension
$\gamma_m$, which effectively rescales the current masses, and the presence
of the conformal anomaly contribution.
We can thus separate $M_B$ into a ``current'' contribution
$M_B^C$ and a
dynamical contribution $M_B^D$; $M_B=M_B^C+M_B^D$ where
\eqn\rgisep
{M_B^C=\langle B|\sferm |B\rangle ,
 \qquad M_B^D=\langle B|\sglue + \gamma _m\sferm |B\rangle.}
Notice that this is a scale invariant separation, since
both $M_B^C$ and $M_B^D$ are renormalization group invariant\col.
Accordingly,
we can separate the mass splittings into current and dynamical
contributions by defining $M_1^C$, $M_2^C$ and $M_1^D$, $M_2^D$ in
analogy with \bmass , with $M_1=M_1^C+M_1^D$ and $M_2=M_2^C+M_2^D$.
The same algebra that led to \masssplit\ also gives $\matele{N}{{\cal
H}_8}{N}=M_1^C-\half M_2^C$, so that
\eqn\newoc
{\Delta M\equiv\matele{N}{{\cal O}_8}{N} -\matele{N}{{\cal H}_8}{N}
=M_1^D-\half M_2^D.}

In the conventional argument \chli, $M_B^D$ is neglected with the
result that $\Delta M$ vanishes, and the sigma term is
given by eqns \hamoc\ and \masssplit. This is presumably done on the
grounds that the term in \rgisep\ proportional to the mass anomalous dimension
$\gamma_m$ is small enough to be ignored, whereas that containing the
isosinglet gluon operator $\sglue$ can be assumed to make no contribution
to mass splittings. We will now show that this
naive assumption is incorrect; away from the chiral limit
$M_1^D$ is actually rather large, and with opposite sign to
$M_1^C$, so that the magnitude of $\matele{N}{{\cal H}_8}{N}$, and thus
of $\sigma_8$, is significantly increased.\foot{Of course a small
nonzero value of $M_1^D$ is present due to the term proportional to the
mass anomalous dimension. This is actually as
large as 30\% of $M_1^C$, since at the nucleon scale
$\gamma_m(1\;\Gev)= 0.27$ (to two loop order)
\ref\Tar{R.~Tarrach, {\it Nucl. Phys.} \vpy{B196}{45}{1982}.}.
This contribution
has (obviously) the same sign as $M_1^C$ and accordingly if taken into account
(while neglecting anything else) would make the disagreement between
the value of $\sigma$ and that of $\sigma_8$ worse, because in eq.\hamoc\
$\sigma_8$ would be replaced by its rescaled value $\sigma_8^R={1\over
1+\gamma_m}\sigma_8$. However including such a contribution while neglecting
that of the gluonic operator is a rather dubious procedure,
since  the separation of $M_B^D$ into a ``quark'' and ``gluon''
piece is necessarily dependent on
the renormalization scale.}

\subsec{Estimation of the Dynamical Mass Splitting}

To this purpose, it is convenient to rewrite the
matrix element on the \lhs\ of eq.\ward\ in terms of one-particle
irreducible physical couplings, analogously to what is done in the
pseudoscalar case in order to derive the isosinglet Goldberger-Treiman
relation\ref\vene{G.~M.~Shore and G.~Veneziano, {\it Phys. Lett.}
\vpy{B244}{75}{1990};\hfill\break {\it Nucl. Phys.} \vpy{B381}{3}{1992}.}.
This can be done by
defining a connected generating functional, $W(S^D_{\mu},S_A)$ where
$S^D_\mu$ is the source for the dilation current $j^\mu_D$, and $S_A$ are
sources for the set of fields
$\Phi_A\equiv(B,\,\bar B,\;\phi_i\equiv{\bar\psi}_i{\psi}_i,\;Q\equiv\sglue )$
(which includes the baryons, the scalar quark condensates,
and the gluon condensate respectively).
Also, as in ref.\vene\ we define the Zumino effective action
$\Gamma(S^D_{\mu},\Phi^{cl}_A)$ by Legendre transformation
of $W(S^D_{\mu},S_A)$ with respect to the fields $\Phi_A$
(but not $j_D^{\mu}$),
so that $\Gamma(S^D_{\mu},\Phi^{cl}_A)$
generates diagrams which are one-particle irreducible with
respect to $\Phi_A^{cl}$.

Rewriting the Ward identity \ward\ in terms of the generating
functionals $W$ and $\Gamma$ it can then be shown \got\ that
\eqnn\canca\eqnn\cancag
$$
\eqalignno{M_B&=(1+\gamma_m)
\sum_i m_i{\delta^3 W\over \delta S_i\delta S_B\delta S_{\bar B}}+
 {\delta^3 W\over \delta S_Q\delta S_B\delta S_{\bar B}}&\canca\cr
&=-\Delta_{\bar{B}\bar{B}}\sum_i
{\delta^3\Gamma\over \delta\phi_i^{cl}\delta B\delta\bar B}
\langle\phi_i^{cl}\rangle \Delta_{BB},&\cancag\cr}
$$
where $\Delta_{BB}$ ($\Delta_{\bar{B}\bar{B}}$) is the baryon
(antibaryon) propagator.
It follows that the \OPR\ contributions to the matrix elements of
the gluon and quark  condensates in eq.\canca\ must cancel against each other
in order
to yield the \OPI\ coupling of eq.\cancag, namely
\eqn\cancb
{\langle B|\sglue |B\rangle^{\rm opr}+\left(1+\gamma_m\right)
\langle B|\sferm |B\rangle^{\rm opr}=0,}
or, in the notation of eq.\rgisep,
\eqn\cancc
{(M_B^C)^{\rm opr}+(M_B^D)^{\rm opr}=0.}
This shows immediately that it is quite unreasonable to assume
$M_B^D$ to be isosinglet given that (just as
in the pseudoscalar sector, where similar results hold
\vene,\ref\gtw{D.~G.~Gross,
S.~Treiman and F.~Wilczek, {\it Phys. Rev.} \vpy{D19}{2188}{1977}.})
one would expect the \OPR\ contributions to $M_B^C$
to have a significant flavor--nonsinglet component.

More specifically,
the \OPR\ matrix elements are presumably dominated by
diagrams where the various operators couple directly to a
meson state with the appropriate quantum numbers,
which then in turn couples irreducibly to the baryon;
\eqn\cou
{(M_B^C)^{\rm opr}={\langle B|\sferm |B\rangle }^{\rm opr}\approx
\sum_{a}\langle
0|\sferm |\phi_a\rangle {1\over
{m^2_{\phi_a}}}\langle\phi_{a}|\bar B B\rangle ,}
where $|\phi_{a}\rangle $ are the scalar meson states, and the sum
over states implicitly includes an integration over momenta.
We can roughly estimate the order of magnitude of this correction by
assuming that it is dominated by the exchange of
two ideally mixed scalar mesons
$\phi_l$ and $\phi_s$, which are respectively an isosinglet
pure $u$ and $d$ state, and a pure $s$ state.
Introducing meson-to-vacuum coupling constants $g_\phi$ and
meson-baryon couplings $g_{\phi \bar B B}$, normalized according to
\eqn\canorm{
\eqalign{&\langle0|\half\left(\bar u u +\bar d d \right)
|\phi_l\rangle=g_{\phi_l} m^2_{\phi_l},\quad
\langle0|\bar s s |\phi_s\rangle= g_{\phi_s} m^2_{\phi_s},\cr
&\langle\phi_a|\bar{B}B\rangle
=(2\pi)^4\delta^4(p_{\phi}-p_B+p_{\bar{B}})g_{\phi_a\bar B B },
\cr
}}
we find
\eqn\corrns{\eqalign{
(M_B^D)^{\rm opr}=-(M_B^C)^{\rm opr}&\approx -2\bar{m}g_{\phi_l}
g_{\phi_l\bar B B}-
m_sg_{\phi_s} g_{\phi_s\bar B B }\cr
&\simeq -(4\bar{m}+m_s)gg'+(m_s-2\bar{m})\matele{B}{Y}{B}gg',\cr}}
where in the second line we further assume the couplings to be given
approximately by the additive quark model with SU(3) symmetry,
so that $g=g_{\phi_l}\simeq g_{\phi_s}$ while $g_{\phi_l\bar{B}B}
\simeq\matele{B}{Y+2}{B}g'$ and $g_{\phi_s\bar{B}B}
\approx\matele{B}{1-Y}{B}g'$.

To estimate $(M_B^D)^{\rm opi}$ we return to Weinberg's
determination \Weinberg\ of the quark masses, which is performed assuming that
the baryon matrix elements of $\bar\psi\psi$ are proportional to
the quark number, and
is phenomenologically accurate to about 20\%. Because all contributions
to $\bar\psi\psi$ such that this identification is correct come from \OPI\
diagrams\pbppap\got , the successfulness of this approach may now
be understood as a consequence of the
cancellation eq.\cancc, thus implying that $(M_B^D)^{\rm opi}$ is approximately
flavor singlet\foot{A discussion of the dynamical reason
for this within the light-cone parton model is given in ref.\pbppap .}.

In view of this, we may conclude that the bulk of the flavor nonsinglet
component of $M_B^D$ is provided by $(M_B^D)^{\rm opr}$, and thus that
$M_1^D\approx (m_s-2\bar{m})gg'$ while $M_2^D\approx 0$. Thus we estimate
$\Delta M$ in eqn. \newoc\ to be very roughly of
order $200 \; \Mev $ if the couplings are of order unity, which is of
the same order of magnitude, and the same sign, as the matrix element
$\matele{N}{{\cal O}_8}{N}$ deduced from the mass splittings
\masssplit. Consequently the theoretical estimate of $\sigma_8$
is approximately doubled, to around
$50\;\Mev $, which is quite compatible with the identification of
$\sigma_8$ and $\sigma$ without the necessity for any violation of
the Zweig rule.

Whereas this estimate of the size of $\Delta M$ relies on the crude
pole-dominance approximation \cou\ used to deduce eq.\corrns, and on the
neglect of $(M_B^D)^{\rm opi}$, the failure of the identification of
the mass splitting operator ${\cal O}_8$ with ${\cal H}_8$ may be inferred from
the exact results \ward, \newoc\ and \cancc. Clearly
all of the usual consequences of the assumption that baryon mass
splittings are given by matrix elements of an octet
operator ${\cal O}_8$, such as the Gell-Mann--Okubo mass
formula, are intact; it is only the identification of ${\cal H}_8$ with
that operator which is modified. The same applies to results derived assuming
that an additive quark model picture applies to the matrix elements
of $\bar\psi\psi$, such as Weinberg's determination of current quark masses
\Weinberg, since these only test the identification of baryon matrix
elements of ${\cal O}_8$ with
$(M_B^C)^{\rm opi}$ which is correct due to the cancellation eq.\cancc.
The sigma term puzzle is thus resolved by the
observation that the gluon condensate provides a positive
contribution to the parameter $M_1$ (which is negative) in eq.\bmass;
the quark's contribution to the splittings is necessarily
underestimated if this is not taken into account.

\newsec{Computation of Dynamical Quark Masses}

How can one test quantitatively the explanation of the sigma term puzzle
proposed in the previous section? Because the gluon condensate
provides the bulk of the baryons' masses, the SU(3)-dependence of its matrix
elements correspond to a similar SU(3) dependence of the dynamical
contribution to constituent quark masses. In particular,
the dynamically generated mass should display a strong dependence on
the current mass, anticorrelated to it.

\subsec{The Quark Self Energy}

We can try to find the flavor dependence
of the dynamically generated mass by studying the
quark self energy, which we may compute by solving the quark \SDE
\nref\LAP{K.~Lane, {\it Phys. Rev.} \vpy{D10}{2605}{1974}\semi
T.~Appelquist and E.~Poggio, {\it Phys. Rev.} \vpy{D10}{3280}{1974}.}
\nref\pagels{H.~Pagels, {\it Phys. Rev.} \vpy{D19}{3080}{1979}.}
\nref\MirHig{V.~A.~Miransky, {\it Sov. Jour. Nucl. Phys.} \vpy{38}{280}
{1983}\semi K.~Higashijima, {\it Phys.Rev.} \vpy{D29}{1228}{1984}.
}\LAP--\MirHig;
\eqn\SDEusual
{S^{-1}(p)=Z_2\;S_0^{-1}(p) - Z_1\;g^2 \int \frac{d^4 k}{(2\pi )^4}
\lambda ^a\gamma _\mu D^{ab}_{\mu\nu }(p-k)S(k)\Gamma
^b_{\nu}(p,k),}
where $S(p)\equiv Z(p^2 )/(\pslash +\Sigma(p^2 ))$ is the full quark
propagator, $\Sigma (p^2 )$ is the quark self energy,
$S_0(p)\equiv (\pslash+m_0)^{-1}$ is the
bare (or ``current'') quark propagator,
$D^{ab}_{\mu\nu}(p-k)$ the full gluon propagator, and
$\Gamma ^b_{\nu}(p,k)$ the full quark-gluon vertex function.
$Z_1$, $Z_2$ and $Z_3$ are the usual renormalization
factors for the vertex, quark and gluon fields, respectively.
The generators $\lambda _a$ of the color group $SU(3)$ are normalized as
$\hbox{tr}(\lambda _a\lambda _b)=\half\delta _{ab}$, and henceforth we
will work throughout
in Euclidean space.

As it stands eq.\SDEusual\ is not consistent with the \RGE s for $S$,
$D$ and $\Gamma $, so it must be renormalization group improved
\ref\BTSDE{R.D.~Ball and J.~Tigg, Oxford preprint OUTP-92-34P.}; this
results in the replacement of the coupling constant with a running
coupling $\bar{g}^2(p,k)$ inside the integral, where
$\bar{g}(k,p)=\bar{g}(p,k)$ and $\bar{g}(k,p)\sim {g}(k^2 )$
when $k^2 \gg p^2$, $g(k^2 )$ being the usual running coupling
\ref\runcoup{C.~H.~Llewellyn~Smith, {\it Acta Phys. Austr.} \vpy{Suppl
XIX}{331}{1978}\semi Yu.~L.~Dokshitzer, D.~I.~Diakonov and
S.~I.~Troyan, {\it Phys. Rep.} \vpy{58}{269}{1980}.}\MirHig .

Let us now assume (neglecting ghost contributions) that
$\Gamma^a_{\nu} =\lambda ^a\Gamma _{\nu}$, where $\Gamma _{\nu}$
satisfies the Ward--Takahashi identity
\eqn\WTI
{(p-k)_{\nu}\Gamma _{\nu}(p,k)=S^{-1}(p)-S^{-1}(k).}
Besides ensuring that $Z_1=Z_2$, eq.\WTI\ may be used to find a consistent
expression for the vertex in terms of the quark propagator. A suitable
solution was given long ago by Landau
\ref\LAK{L.~D.~Landau,A.~Abrikosov and I.~Khalatnikov, {\it Nuovo Cim.
Suppl.} \vpy{3}{80}{1956}.}:
\eqn\Landauvert
{\Gamma _{\mu}(p,k)=\frac{(p-k)_{\mu}}{(p-k)^2}\left(S^{-1}(p)-S^{-1}(k)\right)
+ T_{\mu\nu}(p-k)
\gamma _{\nu}\bar{Z}(p,k)^{-1},}
where $T_{\mu\nu}^{p}\equiv (\delta_{\mu\nu}-p_{\mu}p_{\nu}/p^2)$,
$\bar{Z}(p,k)=\bar{Z}(k,p)$ and $\bar{Z}(p,k)\sim Z(k^2 )$ for
$k^2\gg p^2$.
Even though this ansatz has a kinematic singularity in the infrared
(i.e. as $p\to k$), it is sufficient for
our purposes since it guarantees multiplicative renormalizability, and
thus will give a quark self energy which has the correct gauge
independent asymptotic behaviour at large $p^2$ (see
ref.\BTSDE\ for a more complete discussion).
This confines ambiguities to the infrared region over which we have
little control anyway.
Analogously, the gluon propagator can be taken to have its asymptotic form, up
to a momentum-dependent function $d(p)$
which parametrizes infrared uncertainties;
\eqn\glueprop
{D^{ab}_{\mu\nu}(p)=\delta
^{ab}d(p)\frac{ T_{\mu\nu}(p)+\xi p_{\mu}p_{\nu}/p^2}{p^2},}
where $d(p)\to 1$ as $p^2\to\infty$.

Using the vertex  \Landauvert\  and the propagator \glueprop,
the \SDE\ \SDEusual\ becomes
\eqn\SDEren
{\eqalign{S^{-1}(p)= Z_2&S_0^{-1}(p)+ 3C_2Z_2\int
\frac{d^4k}{(2\pi )^4}\frac{\bar{g}^2(p,k)d(p-k)}{(p-k)^2}\cr
&\times\left[T_{\mu\nu}(p-k)\gamma_{\mu}S(k)\gamma _{\nu}
\bar{Z}(p,k)^{-1}+\xi\frac{(\pslash-\kslash)}{(p-k)^2}
\left(S(k)S^{-1}(p)-1\right)\right],\cr}}
where the colour factor $C_2=\frac{N_c^2-1}{2N_c}=\frac{4}{3}$.

Eq.\SDEren\ can be further simplified by assuming
that the a priori unknown factors $\bar{g}(p,k)$, $d(p-k)$,
and $\bar{Z}(p,k)$ can each be approximated by the asymptotic
values they take
when one of their arguments is much larger
then the other, i.e.
$\bar{g}(p,k)^2d(p-k)\simeq g(\hbox{\rm max}(p^2 ,k^2))^2$
and $\bar{Z}(p,k)\simeq Z(\hbox{\rm max}(p^2 ,k^2))$\LAK .
We may then perform the angular integrals analytically to give
\eqn\SDEsimp
{\eqalign{Z^{-1}(p^2)&=Z_2+\frac{\xi Z_2 C_2}{16\pi ^2}\biggl[
\half g^2(p^2)
-\frac{g^2(p^2)}{p^4}\frac{\Sigma(p^2 )}{Z(p^2 )}
\int _0^{p^2} dk^2\;\frac{k^2 Z(k^2 )\Sigma(k^2 )}{(k^2+\Sigma ^2(k^2) )}\cr
&\qquad\qquad\qquad\qquad +Z(p^2)^{-1}\int _{p^2}^{\Lambda ^2}
dk^2\;\frac{g^2(k^2)Z(k^2)}{k^2+\Sigma ^2(k^2)}\biggr],\cr
\Sigma (p^2)&=Z(p^2)Z_2 m_0(\Lambda ) + \frac{3C_2}{16\pi^2}\biggl[
(1+\third\xi )\frac{g^2(p^2)}{p^2}\int_0^{p^2}dk^2\;
\frac{k^2 Z(k^2)\Sigma(k^2)}{k^2+\Sigma ^2(k^2)} \cr
&\qquad+Z(p^2)\int _{p^2}^{\Lambda^2}dk^2\;
\frac{g^2(k^2)\Sigma (k^2)}{k^2+\Sigma ^2(k^2)}
+\third\xi\Sigma (p^2)\int _{p^2}^{\Lambda^2}dk^2\;
\frac{g^2(k^2)Z(k^2)}{k^2+\Sigma ^2(k^2)}\biggr]. \cr}}
where we have introduced an ultraviolet momentum cutoff $\Lambda $.
In the Landau gauge $\xi =0$,
$Z(p^2)=Z_2^{-1}$, and it is not difficult to see that \SDEsimp\
reduces to the same equation as obtained in the
(renormalization group improved) ladder approximation \MirHig.

Eq.\SDEsimp\ is sufficiently simple that it may
be solved numerically
for the quark self energy $\Sigma(p^2)$. All
infrared uncertainties (in the regions $k^2\sim M^2$, $p^2\sim M^2$ and
$(p-k)^2\sim M^2$, where $M^2$ is the ``QCD scale'') have been rather crudely
absorbed into the infrared uncertainty in the form of the running
coupling $g(p^2)$. By taking suitable parameterizations of this
uncertainty we hope to get some feel for the overall uncertainty due
to our ignorance of the physics of the infrared region. We may
assume for example that
\eqn\coupling
{g^2(p^2)=\frac{1}{\beta_0\ln (\delta+p^2/M^2)},}
where $\beta_0 = \frac{11N_c-2N_f}{48\pi ^2}$, and $\delta$ may be
adjusted at will to give various values of the strong coupling at $p^2=M^2$.

\subsec{Current Mass and Dynamical Mass}

After performing such a calculation, it is still necessary to adopt
some procedure to extract the quark ``mass'' from its self energy.
Since we only have a rather approximate form for the self energy in
Euclidean space, there is no question of being able to continue to
Minkowski space to search for a pole in the quark propagator, even if
one believed that such a pole should exist.
We may however estimate the region of $p^2$ in which the self energy
makes its dominant contribution to the mass of an on--shell baryon.
The simplest such estimate is to take the
``pole--mass'' defined as the solution to
$m_{\Sigma }=\Sigma (m_{\Sigma }^2)$: the self energy is taken to yield the
mass when evaluated at the scale set by the mass itself.

We may assess the uncertainty in this estimate by observing
that the baryon mass could be determined, if we
knew its Bethe--Salpeter amplitude, by
calculating an integral over the relative momenta
of its constituent quarks; the effect of these integrals would be to
give the contribution of each quark as some weighted mean of its self
energy. We may attempt to simulate this by smearing the quark's self-energy
with a momentum-dependent ``form factor'' $\rho(p^2)$, and then
defining the
mass $m_{\Sigma }^{\rho}$ as the solution of
\eqn\massav
{m_{\Sigma }^{\rho}=\langle\Sigma\rangle_{\rho}\equiv\int_0^{\infty}dp^2\;
\rho (p^2)\Sigma (p^2),}
where $\rho$ satisfies the normalization condition
$\langle 1\rangle _{\rho}=1$ and the additional condition
$\langle p^2\rangle _{\rho}=(m_{\Sigma}^{\rho})^2$,
which ensures
that the pole--mass definition is reproduced when $\rho$ tends to a Dirac
delta.
A suitable form of $\rho$
could be $\rho (p^2)={\cal N}p^2(p^2+a)^{-n}$, for $n=4,5,6$, say
\foot{Both the quark propagators and the Bethe--Salpeter amplitude
fall as powers when $p^2$ becomes large.}, with
$\cal N$ and $a$ determined by the two conditions.

The masses $m_{\Sigma}$ determined in this way may be thought
of as ``constituent''
masses; before we can use them to estimate $\Delta M$ we must still
separate out the renormalized ``current'' mass $m_C$ from the
dynamical mass $m_D$ which arises nonperturbatively to break chiral
symmetry even when $m_C$ is small. This can be done by observing that
the amount of explicit symmetry breaking may be extracted from the
asymptotic behaviour at large $p^2$ of the self energy \LAP
\eqn\sigmaasymp
{\Sigma(p^2)\p2inf m\ln ^{-\lambda }(p^2/M^2)(1+O(g(p^2)^2),}
where $\lambda\equiv\gamma _m^0/\beta_0 = \frac{9(N_c^2-1)}{2N_c(11N_c-2N_f)}
=\frac{4}{9}$ when $N_c=N_f=3$, and $\gamma _m^0$ is the leading coefficient
in the perturbative expansion of the mass anomalous dimension. The
renormalization group invariant mass parameter $m$ is related to the
running current mass in the usual way:
\eqn\rmass
{m_C(m,\mu )\twiddles{\mu} m\ln ^{-\lambda}(\mu^2/M^2),}
where $\mu$ is the renormalization scale.

By varying the bare mass $m_0(\Lambda )$ we obtain thus a
family of solutions to \SDEsimp, each of which corresponds to a
different (unique) value of $m$. Calculating the constituent mass
$m_{\Sigma}$ for each of these solutions gives us $m_{\Sigma}$ as
a function of $m$. The asymptotic behaviour \sigmaasymp\ in
combination with the definition of $m_{\Sigma}$ tells us immediately that
\eqn\mbarasymp
{m_{\Sigma}(m)\twiddles{m}m
\ln ^{-\lambda }(m^2/M^2)(1+O(g(m^2)^2)).}

To obtain the dynamical mass $m_D(m)$ we should now subtract from
$m_{\Sigma}(m)$ the current mass $m_C$ eq.\rmass,
evaluated at the scale of the constituent mass:
\eqn\dynmass
{m_D(m)= m_{\Sigma}(m)-m_C(m,m_\Sigma(m)).}
Due to the asymptotic behaviors of $m_\Sigma$, eq.\mbarasymp, and $m_C$,
eq.\rmass, the dynamical mass defined by eq.\dynmass\
vanishes for large $m$, as it ought to
(the current mass and the constituent mass coincide for large values of $m$).

It is still necessary however to further
specify the form of $m_C$ in the infrared\foot{Even if we could compute
$m_C(m,\mu )$ reliably for small $\mu$, we would still require extra conditions
to fix the (probably rather large) scheme dependence.}.
Clearly $m_{\Sigma}(0)=m_D(0)$, so $m_C(0,m_D(0))=0$.
Furthermore we may use Weinberg's definition \Weinberg\
of the constituent quark mass, i.e., assume that in the
chiral limit the dependence of $m_D$ on $m$ is dominated
by the linearized dependence of $m_C$ on $m$ (which, recalling eq.\rgisep,
is essentially the dependence of $M^C_B$ on the quark content of the given
state).
In other words we assume that for small $m$
\eqn\Weinmassa
{m_{\Sigma}(m)\simeq m_D(0)+m_C(m,m_{\Sigma}(m)),}
and, identifying the linear terms in a Taylor expansion in $m$,
\eqn\Weinmassb
{\frac{dm_C}{dm}\bigg\vert_{m=0}=\frac{dm_{\Sigma }}{dm}\bigg\vert
_{m=0}.}
Note that this condition  may  be enforced only at one
particular $m$ since $m_{\Sigma}(m)$ is nonlinear.
In practice we choose a smooth
interpolating function to parameterize $m_C$ for all $m$, such as, for example
(compare eq.\coupling )
\eqn\runningmass
{m_C(m,m_{\Sigma}(m) )=m\ln ^{-\lambda }(\epsilon +m^2/M^2),}
where the parameter $\epsilon $ is chosen to ensure the satisfaction
of \Weinmassb.

         The dynamical mass $m_D(m)$ is then  a smooth function,
flat at the origin, tending gradually to zero on a scale of
$m_D(0)\approx 300\;\Mev$.
It is this mass which should be identified with the dynamical
contribution $M_B^D$ (see \rgisep ) at the constituent quark level,
since the ``current'' quark contribution $M_B^C$ is
clearly already accounted for by
$m_C$. Thus, referring back to \bmass\ and \newoc\ we see that
assuming a linear mass formula
\eqn\Delm
{\Delta M\simeq M_1^D\simeq m_D(0)-m_D(m_s),}
where $m_s$ is the value of $m$ appropriate for the strange quark
mass; taking the running mass
$m_C(m_s,1\;\Gev)=175\pm 50\;\Mev$ \mas\ we find
$m_s=200\pm 50\;\Mev $ \mas, which is of the same order of
magnitude as $m_D(0)$. From the qualitative behaviour of $m_D(m)$
we thus expect $\Delta M$ to be loosely of order $150 \Mev $, in
rough agreement with the estimate of section 2.2.

\subsec{Numerical Results}

We may now compute $\Delta M$ eq.\Delm\ by
solving the \SDE\ in the approximate form \SDEsimp ,
with $N_c=N_f=3$ and, for example, the ansatz \coupling , and
extracting $m_D(m)$ according to \massav, \dynmass\ and \runningmass .
The results of such calculations with
$\alpha _s(0)\equiv (4\pi\beta_0\ln\delta)^{-1}=2.5$ and the
gauge parameter $\xi =0$ are displayed in
fig.2, for both the ``pole'' definition and the various averaging
definitions \massav\ of the constituent mass; it can be seen that
$m_D(m)$ depends very little on which definition we adopt, so all
remaining calculations are done with the simpler ``pole'' definition.

The variation with gauge parameter $\xi$ is explored in fig.3 --- again
we find that for $|\xi|\lsim 1$ $m_D$ is approximately independent of
the choice of gauge, as expected since the Landau vertex \Landauvert\
satisfies the Ward--Takahashi identity \WTI . Of course for larger
values of $\xi$ there is still some gauge dependence, resulting from
the infrared uncertainties. This could presumably only be tamed
by using a vertex which satisfied the full
Slavnov--Taylor identity, and thus including ghost contributions which are
significant in the infrared, but this is not possible with the present level
of expertise\nref\BCCP{J.~S.~Ball and T.-W.~Chiu, {\it Phys. Rev.}
\vpy{D22}{2542}{1980}
\semi D.~C.~Curtis and M.~R.~Pennington, {\it Phys. Rev.}
\vpy{D42}{4165}{1990}.}\foot{It is however possible to do
computations with more
complicated vertices \BCCP\ which satisfy the Ward--Takahashi identity \WTI\
but are however free of the infrared kinematic singularity which plagues
\Landauvert . The results are however little different to those with
the simpler Landau vertex.}.

However the dominant uncertainty comes not from gauge dependence but
from our general ignorance of the infrared. If we vary the strong
coupling in the infrared by choosing different values for
$\delta$ in \coupling , we can obtain wide
variations in $m_D(m)$; displayed in fig.4 are curves for
$\delta = 2,1.5,1.2,1.01$ corresponding to
$\alpha _s(0)= 0.5,1,2.5,50$. Clearly as $\alpha _s(0)$ is increased the
dynamical mass becomes firmer, and consequently falls off less rapidly
as $m$ is increased; $\Delta M$ is thus reduced. To
check this observation we also computed curves for couplings
containing various infrared singularities, for example $\delta ^4(p)$
or $1/(p^2)^a$ with $0\leq a<1$ (for $a=1$ the \SDE\ is infrared
divergent) which confirm this expectations: the stronger the
singularity, the firmer the dynamical mass.

We conclude that it is
not possible to compute $\Delta M$ without further (highly nontrivial)
information on the infrared structure of QCD --- all we can do is
estimate $\Delta M$ to lie in the range $50\;\Mev\lsim\Delta M\lsim
250\Mev $
which translates (using \newoc ) into the estimate
\eqn\sigmaest
{30\;\Mev\lsim\sigma _8\lsim 60\;\Mev .}
in perfectly satisfactory agreement with the value of $\sigma $
inferred from experiment (and with the estimate presented in section 2.2).

Amusingly, we also have sufficient information to compute $\sigma $
using the so-called Feynman--Hellman formula at the constituent quark level,
which amounts to the (rather reasonable) assumption that $m_C$ is
approximately linear in the small light flavor average mass $\bar m$:
\eqn\FHquark
{\sigma\simeq 3\bar{m}\frac{\partial m_{\Sigma }}{\partial \bar{m}}
= 3m_C(\bar{m},m_D(0))}
where in the last step we used eqns.\Weinmassa, \Weinmassb,
and we may compute $m_C(\bar{m},m_D(0))$ through eq.\runningmass\ from the
known value of the running current mass at the nucleon scale.
Taking $m_C(\bar{m},1\;\Gev)=5\pm 3\;\Mev $ \mas,\DHW,
we find $\sigma_8 = 40\pm 20\pm 15\Mev $, the first error being due to
the uncertainty in $m_C(\bar{m},1\;\Gev)$, and the second due to the
uncertainty in $\epsilon $, which we estimate by varying
$\alpha _s(0)$ from $1$ to $50$. Since the former is rather large, and the
latter probably underestimated, this is clearly a less reliable
result than \sigmaest , but nonetheless it provides a useful
independent check on the consistency of the calculation.

The calculations presented here are necessarily extremely uncertain,
not only due to our implicit use of a constituent quark picture of the baryon
(instead of a linear mass formula, we could have used a more
sophisticated one (see for example \ref\IsKa{A.~De Rujula, H.~Georgi
and S.L.~Glashow, {\it Phys. Rev.} \vpy{D12}{147}{1975}\semi
N.~Isgur and G.~Karl, {\it Phys. Rev.} \vpy{D18}{4187}{1978};
\vpy{D19}{2653}{1979}.}) in which $M_2^D$ would also be nonzero), but also to
uncertainties in the separation of the dynamical contribution
from the ``current'' quark contribution, in the form of the
quark--gluon vertex function and, most importantly,
our complete ignorance of the behaviour of the gluon propagator
in the infrared. Note that the possible effects of quark loops, both
light and heavy, are only a part of this uncertainty, in that while
they certainly contribute to the gluon propagator in the infrared,
there is no reason to believe that they dominate it.

A more
accurate computation of the sigma term than that presented here
would thus require much more powerful techniques (for example
lattice studies), because of its sensitivity to the infrared
dynamics of chiral symmetry breakdown. Certainly nothing more can be learnt
from models such as that of Nambu--Jona-Lasinio, or the chiral quark
model, since these necessarily trivialize this dynamical structure by
assuming a constant quark self energy.

\newsec{Conclusion}

In this paper we have shown that the traditional argument relating the
sigma term to baryon mass splittings \chli\ is incorrect, due to
the neglect of the dynamical contribution to such mass splittings from
the trace anomaly. This contribution may be viewed as a nonsinglet gluonic
contribution to constituent masses, and as such it does not spoil all the
usual quark model results, although it does contradict the naive assumption
that mass splittings at the constituent and current level are the same.

We have further attempted a computation of this
contribution through the solution of the Schwinger-Dyson equation
for the quark self energy, and find that it not only has the right sign, but is
also quite large enough to agree with the experimental determination.
However since it is rather sensitive to the infrared dynamics, it
seems to us to be rather difficult to obtain a more precise
determination, and thus a firm bound on the amount of Zweig rule
violation, without a better understanding of the
details of the mechanism responsible for the dynamical breaking of
chiral symmetry. What we can say, however, is that without such an
understanding the experimental value of the sigma term cannot be used
as evidence of strange quarks in the nucleon.

Rather than providing evidence for the breakdown of the naive
quark model, the sigma term appears thus to
provide nontrivial information on the dynamics of QCD in the infrared.
In view of this, both a better
experimental determination of the sigma term, and a better theoretical
understanding of nonsinglet contributions to quark masses would be highly
desirable.

\bigskip
\noindent{\bf Acknowledgement:} RDB would like to thank the Royal
Society for their financial support; JT thanks the SERC. SF thanks
M.~Bos, V.~de~Alfaro
and E.~Predazzi for discussions.

\vfill
\eject
\listrefs
\vfill
\eject
\centerline{\bf Figure Captions}
\bigskip
\item{[Fig.~1]} The \SDE .
\item{[Fig.~2]} The dynamical mass $m_D(m)$ vs. the current mass $m$,
both in units of $m_D(0)$ (solid); also $m_D^n(m)$ (defined in
eqn.\massav ) for n=4,5,6 (long dashes, short dashes, dots respectively).
\item{[Fig.~3]} As fig. 2, but for the gauge parameter
$\xi=0,1,2,-1,-2$ (solid, long dashes, short dashes, dots, dots and
dashes respectively).
\item{[Fig.~4]} As fig. 2, but for $\alpha_s(0)=0.5,1,2.5,50$ (dots,
short dashes, solid, long dashes respectively).
\vfill
\eject
\bye